\documentclass{article}

\usepackage{arxiv}

\usepackage[utf8]{inputenc} 
\usepackage[T1]{fontenc}    
\usepackage{hyperref}       
\usepackage{url}            
\usepackage{booktabs}       
\usepackage{amsfonts}       
\usepackage{nicefrac}       
\usepackage{microtype}      
\usepackage{lipsum}		
\usepackage{graphicx}
\usepackage{natbib}
\usepackage{doi}
\usepackage{makecell}

\title{NOTE: Solution for KDD-Cup 2021 WikiKG90M-LSC}

\author{ Weiyue Su \\
	Team PGL\\
	Baidu Inc., Shenzhen\\
	China \\
	\texttt{suweiyue@baidu.com} \\
	\And
    Zeyang Fang \\
	Team PGL\\
	Baidu Inc., Shenzhen\\
	China \\
	\texttt{fangzeyang@baidu.com} \\
		\And
    Hui Zhong \\
	Team PGL\\
	Baidu Inc., Shenzhen\\
	China \\
	\texttt{zhonghui03@baidu.com} \\
		\And
    Huijuan Wang \\
	Team PGL\\
	Baidu Inc., Shenzhen\\
	China \\
	\texttt{wanghuijuan03@baidu.com} \\
		\And
    Siming Dai \\
	Team PGL\\
	Baidu Inc., Shenzhen\\
	China \\
	\texttt{daisiming@baidu.com} \\
		\And
    Zhengjie Huang \\
	Team PGL\\
	Baidu Inc., Shenzhen\\
	China \\
	\texttt{huangzhengjie@baidu.com} \\
		\And
    Yunsheng Shi \\
	Team PGL\\
	Baidu Inc., Shenzhen\\
	China \\
	\texttt{shiyunsheng01@baidu.com} \\
		\And
    Shikun Feng \\
	Team PGL\\
	Baidu Inc., Shenzhen\\
	China \\
	\texttt{fengshikun01@baidu.com} \\
		\And
	Zeyu Chen \\
	Team PGL\\
	Baidu Inc., Shenzhen\\
	China \\
	\texttt{chenzeyu01@baidu.com} 
}

\renewcommand{\headeright}{Technical Report}
\renewcommand{\undertitle}{Technical Report}
\renewcommand{\shorttitle}{PGL at KDD Cup 2021 WikiKG90M-LSC Task}

\hypersetup{
pdftitle={A template for the arxiv style},
pdfsubject={q-bio.NC, q-bio.QM},
pdfauthor={Weiyue Su, Zeyang fang, Hui Zhong, Huijuan Wang, Siming Dai, Zhengjie Huang, Yunsheng Shi, Shikun Feng, Chen Zeyu},
pdfkeywords={First keyword, Second keyword, More},
}

\begin{document}
\maketitle

\begin{abstract}
 WikiKG90M in KDD Cup 2021 is a large encyclopedic knowledge graph, which could benefit various downstream applications such as question answering and recommender systems. Participants are invited to complete the knowledge graph by predicting missing triplets. Recent representation learning methods have achieved great success on standard datasets like FB15k-237. Thus, we train the advance algorithms in different domains to learn the triplets, including OTE, QuatE, RotatE and TransE. Significantly, we modified OTE into 
NOTE (short for Norm-OTE) for better performace. Besides, we use both the DeepWalk and the post-smoothing technique to capture the graph structure for supplementation. In addition to the representations, we also use various statistical probabilities among the head entities, the relations and the tail entities for the final prediction. Experimental results show that the ensemble of state-of-the-art representation learning methods could draw on each other's strength. And we develop feature engineering from validation candidates for further improvements.
 Please note that we apply the same strategy on the test set for final inference. And these features may not be practical in the real world when considering ranking against all the entities.

\end{abstract}



\section{Introduction}

\begin{figure}
	\centering
	\includegraphics[width=0.5\linewidth]{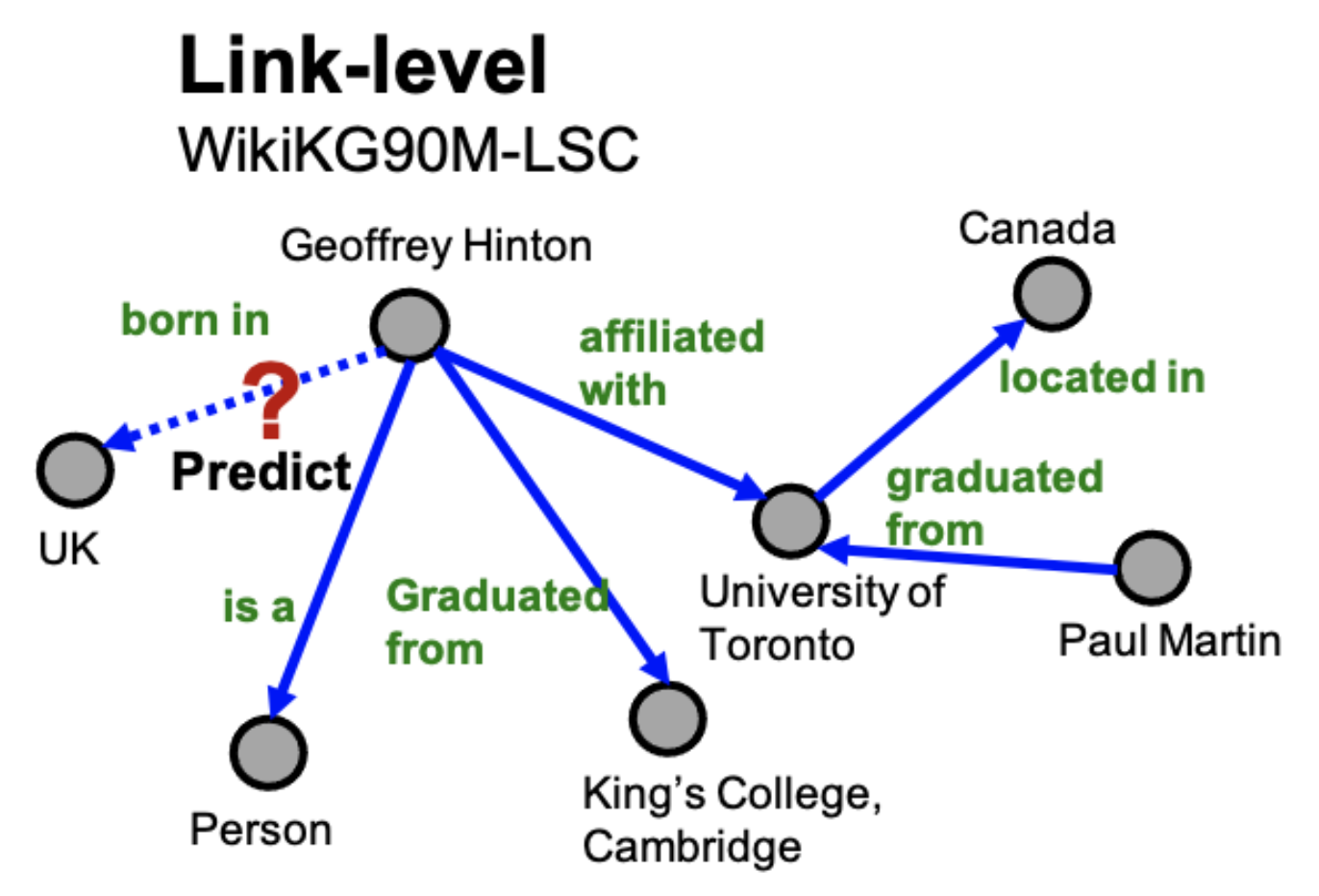}
	\caption{Impute missing triplets in knowledge graphs.\protect\footnotemark[1]}
	\label{fig:wikikg}
\end{figure}

 Knowledge graphs are directed multi-relational graphs about facts, usually expressed in the form of $(h, r, t)$ triplets, where $h$ and $t$ represent head entity and tail entity respectively, and $r$ is the relation between head entity and tail entity, e.g. (\textit{Geoffrey Hinton, citizen of, Canada}). Large encyclopedic knowledge graphs, like Wikidata \citep{wikidata} and Freebase \citep{freebase}, can provide rich structured information about entities and benefit a wide range of applications, such as recommender systems, question answering and information retrieval. However, large knowledge graphs usually face the challenge of incompleteness. For example, 71\% of people in Freebase have no birth place and 75\% have no nationality \citep{dong2014knowledge}. Therefore, predicting these missing facts in a knowledge graph is a crucial task, also named as knowledge graph completion task. We can see Figure  \ref{fig:wikikg} for a clear understanding.
 
 In order to address the issue of knowledge graph completion on large knowledge graphs, the 2021 KDD cup releases the WikiKG90M-LSC Task, which focuses on imputing missing facts in a knowledge graph extracted from the entire Wikidata knowledge base. Our method for this task consists of two stages: On the first stage, we propose an ensemble method for different knowledge embedding methods to build a strong model, in which, knowledge embedding methods, like TransE \citep{transe} and RotatE \citep{rotate}, focus on embedding entities and relations into vectors and then we can use these embeddings to predict missing relations. On the second stage, we adopt several statistical features of the WikiKG90M dataset to help improve the final ensemble model performance. We conducted several experiments on WikiKG90M dataset to demonstrate the superiority of our method, and  our team is currently among the awardees of the WikiKG90M-LSC track of the OGB-LSC, where we achieve 0.9727 MRR result in the final test set. Some features from test candidates are used in final inference, which are not practical when considering ranking against all the entities. And its limitation will be discussed in Section \ref{limit}.
 
 
\section{Methodology}
\footnotetext[1]{https://ogb.stanford.edu/assets/img/ogb\-lsc\-task\-overview.png} 

\subsection{Representation Learning}

\subsubsection{Triplet Embedding} 

The majority of knowledge graph representation algorithms relies on the triplets and they remain good interpretability in graph reasoning. In this competition, we adopt advance algorithms in different domains to encode the entities and relations, including NOTE, QuatE, RotatE and TransE. Considering the properties of each model, we ensemble their score results for the final prediction.

\paragraph{NOTE} \cite{ote} proposes OTE to model the symmetry/antisymmetry, inversion, and composition patterns. It takes the relations as an orthogonal transform in a high dimensional space. Relation matrix is orthogonalized so its inverse matrix can be obtained by simple transposing. The full model can be seen as an ensemble of $K$ OTE models. The score functions are defined as Eq.~\ref{eq:ote_head} and ~\ref{eq:ote_tail}.

\begin{equation}
\label{eq:ote_head}
    d((h,r),t)=\sum_{i=1}^K(\| s_r^h(i)\phi(M_r(i))e_h(i)-e_t(i)\|),
\end{equation}
\begin{equation}
\label{eq:ote_tail}
    d(h,(r,t))=\sum_{i=1}^K(\| s_r^t(i)\phi(M_r(i))^Te_t(i)-e_h(i)\|),
\end{equation}

where $s_r^h(i)=diag({exp}(s_r(i)))$ and $s_r^t(i)=diag({exp}(-s_r(i)))$ are the weights of relation matrix, $\phi$ is the Gram-Schmidt process.

However, though \cite{ote} has scaled the $L_2$ norm of relation embeddings through scalar tensors $s_r(i) \in \mathcal{R}_{d_s}$, the convergence is still unstable in our experiments. The $exp$ operation in $s_r^h(i)$ and $s_r^t(i)$ enlarges values and could lead to this issue. Therefore, we further use $L_2$ norm to regularize the scalar tensors. In this case, the weights of relation matrix are modified into Eq.~\ref{eq:scalar_head} and ~\ref{eq:scalar_tail}. We denote such modified version as NOTE (short for Norm-OTE).

\begin{equation}
\label{eq:scalar_head}
    s_r^h(i)=\frac{diag({exp}(s_r(i)))}{\|diag({exp}(s_r(i)))\|},
\end{equation}

\begin{equation}
\label{eq:scalar_tail}
    s_r^t(i)=\frac{diag({exp}(-s_r(i)))}{\|diag({exp}(-s_r(i)))\|}.
\end{equation}

\paragraph{QuatE} \cite{quate} extend ComplEx \citep{complex} into quaternion space for better geometrical interpretations and more latent inter-dependencies. It proves that the Hamilton product in quaternion space can model both symmetry/antisymmetry and inversion patterns except the composition pattern. 

\paragraph{RotatE} RotatE can be view as an orthogonal transform in complex domain. Each relation is taken as a rotation from the head entity to the tail entity. \cite{rotate} has proved that the Hadamard product is able to model various relation patterns. In theory, it has equivalent model capability as OTE, so we also take it as a basic method.

\paragraph{TransE} \cite{transe} interprets relationships as translations operating on entity embeddings in a real field. Although such assumption is not able to model complex relationships such as 1-N, N-1 and N-N, experiments on different datasets have proved its robustness and it can model composition pattern. Thus, we use results of this model to ensemble in our experiments.

\subsubsection{Graph Context} 

Besides the specific triplets, the structure of sub-graphs comprised of triplets also reflect some semantic information. For example, \textit{Geoffrey Hinton} in Figure ~\ref{fig:wikikg} is connected with \textit{King's College Cambridge} through relation \textit{Graduated from}. We can infer that \textit{Geoffrey Hinton} should be a person and is probable to be born in UK, as \textit{King's College Cambridge} is located in UK. Entities and relations in this sub-graph influence each other. To supplement information of the graph structure, we use representative techniques such as Post-Smoothing \citep{appnp} and  
DeepWalk\citep{deepwalk}.

\paragraph{Relation-based Post Smoothing} We propose a two-stage method to capture the relationships among the entities in sub-graphs. Recent end-to-end models always use graph neural networks as encoder and triplet-based methods as decoder. But in our implementation, we take the opposite approach. In the first stage, we train TransE to encode the entities and relations according to the triplet context. In the second stage, we take the learned representations as entity embeddings. Then we propagate them through the entity adjacent matrix. A hyperparameter $\alpha$ decides the weight of the entity itself while $1-\alpha$ denote weights of its neighborhood. The final updated embeddings are used for prediction. For example, for a given entity u, the final embedding $\mathbf{u}'$ is represented in Eq.~\ref{eq:post-smoothing}.

\begin{equation}
    \label{eq:post-smoothing}
    \mathbf{u}' = \alpha \mathbf{u} + (1-\alpha) \sum_{v \in \mathcal{N}(u)}{f(\mathbf{v}, \mathbf{r})},
\end{equation}
where $\mathbf{u}$ and  $\mathbf{v}$ are the embedding learned in the first stage. $f$ is depending on the knowledge embedding algorithm.

\paragraph{DeepWalk} \cite{deepwalk} propose DeepWalk to learn latent representations of nodes in homogeneous networks. In our solution, the relations in knowledge graph are ignored. We focus on the entity structure and use skip-gram technique to learn the semantic and structural correlations between entities along generated paths.

\subsection{Manual Feature Engineering}

In addition to embedding models, manual feature engineering is also a key part of our work. Since our goal is to predict tail through head and relation, our manual features include two parts, head to tail feature and relation to tail feature. Feature selection needs to be performed after obtaining the features.
\subsubsection{Head to Tail Features}
Head to tail feature is to predict the probability of tail by the current head. We start to walk from the head, and calculate the probability of walking through different paths to reach tail.
For a head, relation and tail triple, it has 6 different walk directions $direct$, including head to tail(HT), head to relation(HR), relation to head(RH), relation to tail(RT), tail to head(TH), tail to relation(TR).  The probability of $e_1$ to $e_2$ in direction $direct$ is as fallow.
\begin{equation}
\label{eq:probability}
P_{direct}(e_1, e_2)=\frac{S_{direct}(e_1, e_2)}{\sum_{e \sim N_{direct}(e_1)}S_{direct}(e_1, e)}
\end{equation}

$S_{direct}(e_1, e_2)$ is the frequency of $e_1$ to $e_2$ in the direction of $direct$. When developing our model, we calculated all features from the training triplets and validation candidates. It is worth nothing that we only calculate the $F_{HT}$ and $F_{RT}$ from test data at the final inference time, without modifying the weight of our developed models. And we apply the same rule strategies developed from validation for final test prediction. That is to say, test data is only touched at the final inference time.


We define 7 manual head to tail feature as follow.

\begin{figure}
	\centering
	\includegraphics[width=0.7\linewidth]{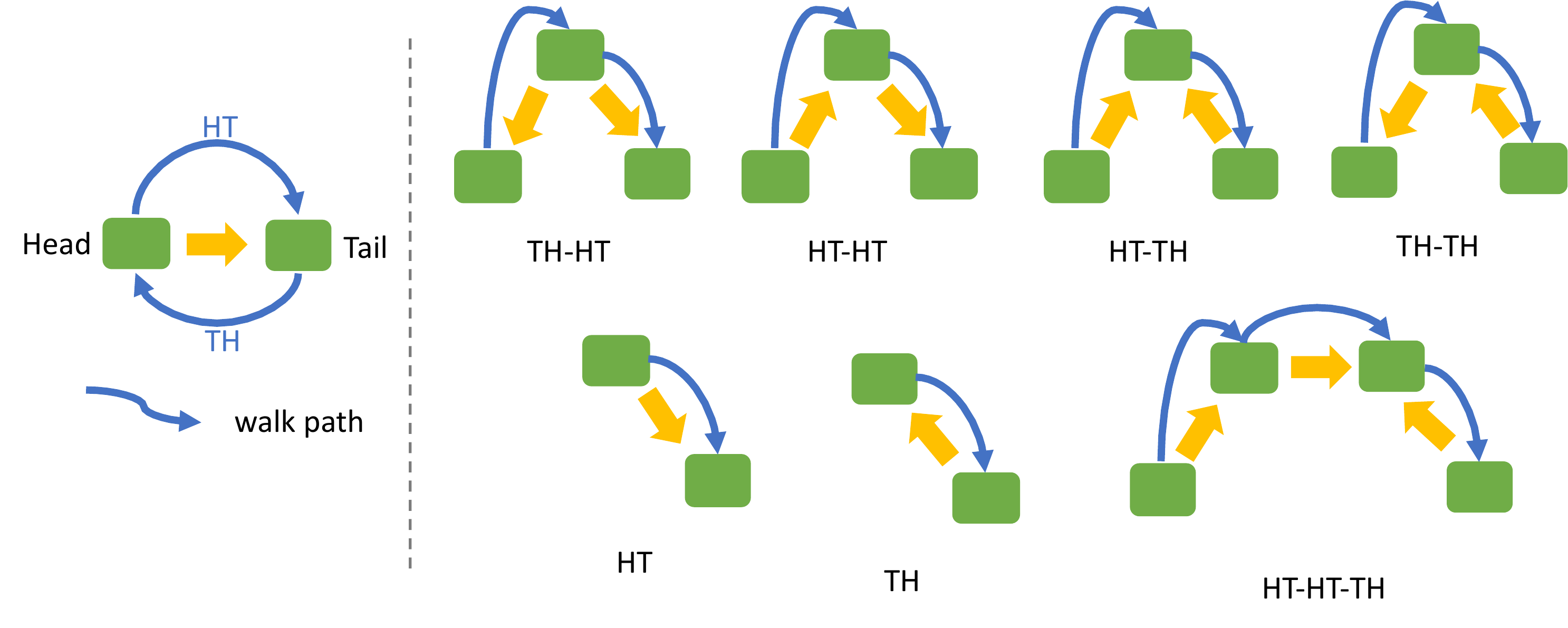}
	\caption{The walk paths of head to tail features. The yellow arrow is a direction of head entity to tail entity.  HT is walk in the direction of head to tail, while TH is in direction of tail to head.}
	\label{fig:htfeature}
\end{figure}

\begin{equation}
\label{eq:featurehtth}
F_{HT}(h, t)= P_{HT}(h, t)
\end{equation}

\begin{equation}
\label{eq:featurehtth}
F_{TH}(h, t)= P_{TH}(h, t)
\end{equation}

\begin{equation}
\label{eq:featurethht}
F_{TH-HT}(h, t)=\sum_{e \sim N_{TH}(h) \cap N_{TH}(t)}P_{TH}(h, e) * P_{HT}(e, t)
\end{equation}

\begin{equation}
\label{eq:featurehtht}
F_{HT-HT}(h, t)=\sum_{e \sim N_{HT}(h) \cap N_{TH}(t)}P_{HT}(h, e) * P_{HT}(e, t)
\end{equation}

\begin{equation}
\label{eq:featurehtth}
F_{HT-TH}(h, t)=\sum_{e \sim N_{HT}(h) \cap N_{HT}(t)}P_{HT}(h, e) * P_{TH}(e, t)
\end{equation}

\begin{equation}
\label{eq:featurethth}
F_{TH-TH}(h, t)=\sum_{e \sim N_{TH}(h) \cap N_{HT}(t)}P_{TH}(h, e) * P_{TH}(e, t)
\end{equation}

\begin{equation}
\label{eq:featurehtth}
F_{HT-HT-TH}(h, t)=\sum_{e_1 \sim N_{HT}(h)} \sum_{e_2 \sim  N_{HT}(t)}P_{HT}(h, e_1) *P_{HT}(e_1, e_2) * P_{TH}(e_2, t)
\end{equation}

\subsubsection{Relation to Tail Features}
We also define 5 manual relation to tail feature as fallow.
\begin{equation}
\label{eq:featurehtth}
F_{RT}(r, t)=P_{RT}(r, t)
\end{equation}

\begin{equation}
\label{eq:featurehtth}
F_{RH}(r, t)=P_{RH}(r, t)
\end{equation}

\begin{equation}
\label{eq:featurehtth}
F_{RT-TR-RT}(r, t)=\sum_{e_1 \sim N_{RT}(r)} \sum_{e_2 \sim  N_{TR}(t)}P_{RT}(r, e_1) *P_{TR}(e_1, e_2) * P_{RT}(e_2, t)
\end{equation}

\begin{equation}
\label{eq:featurehtth}
F_{RH-HR-RT}(r, t)=\sum_{e_1 \sim N_{RH}(r)} \sum_{e_2 \sim  N_{HR}(t)}P_{RH}(r, e_1) *P_{HR}(e_1, e_2) * P_{RT}(e_2, t)
\end{equation}

\begin{equation}
\label{eq:featurehtth}
F_{RT-HR-RT}(r, t)=\sum_{e_1 \sim N_{RT}(r)} \sum_{e_2 \sim  N_{HR}(t)}P_{RT}(r, e_1) *P_{HR}(e_1, e_2) * P_{RT}(e_2, t)
\end{equation}


\subsubsection{Feature Selection}
In order to combine the above-mentioned features and models, we use grid search for feature selection. The grid search will output the weights of the embedding models and manual features.

\subsubsection{Limitation Discussion.}\label{limit}
In practical scene, the tail entities follow the long tail distribution. Finding candidates from all entities involves a very complicated process with strategies like rules and approximate nearest neighbor searching. However, different from our actual application, in this competition, candidates are provided from uniform distribution together with long tail ground truth according to the task description paper from \cite{hu2021ogblsc}. Therefore, it comes a simple strategy to narrow the candidate choices by dropping tail entities with low frequencies counting from candidates. But the long tail relations still depend on the Knowledge Embedding strategies, which can be found in our experimental results. To 
narrow the gap between competition and practical application in the real scene, we suggest that the candidates should not be provided.

\section{Experiments}
\label{sec:others}

\subsection{Experimental Details}

Original WikiKG90M dataset contains three time-stamps: September, October, and November of 2020, for training, validation, and testing, respectively, and only entities and relation types that appear in the earliest September knowledge graph are retained. The default parameter settings all models are batch\_size=1000, learning rate of the mlp (mlp\_lr)=2e-5, learning rate decay step (lrd\_step)=1e-5, learning rate of the embedding (lr)=0.1, gamma=12 and hidden\_size=200. Specifically, ote\_size=20 in NOTE model. For our final submission, we mix the training data and validation data.

\if{}
\subsection{Single Model}
Table \ref{tab:table1} shows the official validation dataset result of single model we use w/o manual features. We can learn that our normalized OTE model NOTE achieves best single model performance.
\begin{table}[h]
    \centering
    \caption{Single Model We Use}
    \begin{tabular}{ccc} 
    \toprule
    \textbf{Model} & \textbf{Without manual features} & \textbf{With manual features}  \\ 
    \midrule
    TransE      & 0.8559      & 0.9537       \\
    RotatE      & 0.8871      & 0.9698       \\
    QuatE       & 0.8846      & 0.9608       \\
    NOTE         & 0.9189      & 0.9748       \\
    \bottomrule
    \end{tabular}
    \label{tab:table1}
\end{table}
\fi

\subsection{Experimental Results}

Table \ref{tab:table2} shows the specific structure of all the models we use for ensemble, and report the final validation and test results of our ensemble method.

\begin{table}[h]

    \centering
	\caption{Experimental Results: "-" means default parameter setting.}
	\scalebox{0.8}{

	\begin{tabular}{ccc}
		\toprule
		\textbf{Model Settings} & \textbf{Main Parameter Settings} & \textbf{Validation Acc.}\\
		\midrule
		NOTE-1 & -  &  0.9189   \\
		NOTE-2     & batch\_size=1200 & 0.9191  \\
		NOTE-3     & lrd\_step=2e5   & 0.9172  \\
		NOTE-4     & lrd\_step=4e4   & 0.9170  \\
		NOTE-5     & mlplr=4e-5  & 0.9179  \\
		NOTE-6     & lr=0.3   & 0.9181 \\
		NOTE-7     & gamma=10   & 0.9160  \\
		NOTE-8     & gamma=14   & 0.9173  \\
    	NOTE-9    & ote\_size=40   & 0.9201  \\
		NOTE-10    & hidden\_size=240  & 0.9182 \\
		TransE   & - & 0.8559  \\
		TransE  & post-smoothing & 0.8800 \\
		RotatE   & -  & 0.8871  \\
		RotatE  & post-smoothing & 0.8940 \\
		QuatE    & - & 0.8846   \\
		DeepWalk & - & 0.6132 \\
				\midrule

		features (from training triples) & - & 0.8457 \\
		features (from validation candidates) & - &  0.7551 \\
		features (from training triples \& validation candidates) & - &  0.8778 \\
		\midrule
		Ensemble the Above Models & w/o features & 0.9435  \\
		Ensemble the Above Models & w/ features (from training triples)& 0.9492  \\
		Ensemble the Above Models  & w/ features (from training triples \& validation candidates) & 0.9797  \\
		\bottomrule
	\end{tabular}
	}
	\label{tab:table2}
	
\end{table}

\section{Conclusion}

In this paper, we present our solution for the final test. First, we ensemble different models for representation learning. Specifically, we propose the NOTE model to make the training process steady.
And the DeepWalk and the post-smoothing technique are used to capture the graph structure information among learned embeddings. 
We also use recent advance models including QuatE, RotatE, TransE.
Second, we detail the manual feature engineering. The selected features are used to adjust the predictions. The experimental results show that our solution achieves excellent performance on the WikiKG90M dataset. For the final submission, we apply the same strategy on the test candidates in final inference, and these features are not practical when considering ranking against all the entities.

\bibliographystyle{unsrtnat}
\bibliography{references}

\begin{thebibliography}{11}
\providecommand{\natexlab}[1]{#1}
\providecommand{\url}[1]{\texttt{#1}}
\expandafter\ifx\csname urlstyle\endcsname\relax
  \providecommand{\doi}[1]{doi: #1}\else
  \providecommand{\doi}{doi: \begingroup \urlstyle{rm}\Url}\fi

\bibitem[Vrandecic and Kr{\"{o}}tzsch(2014)]{wikidata}
Denny Vrandecic and Markus Kr{\"{o}}tzsch.
\newblock Wikidata: a free collaborative knowledgebase.
\newblock \emph{Commun. {ACM}}, 57\penalty0 (10):\penalty0 78--85, 2014.

\bibitem[Bollacker et~al.(2008)Bollacker, Evans, Paritosh, Sturge, and
  Taylor]{freebase}
Kurt~D. Bollacker, Colin Evans, Praveen Paritosh, Tim Sturge, and Jamie Taylor.
\newblock Freebase: a collaboratively created graph database for structuring
  human knowledge.
\newblock In \emph{Proceedings of the {ACM} {SIGMOD} International Conference
  on Management of Data, {SIGMOD} 2008, Vancouver, BC, Canada, June 10-12,
  2008}, pages 1247--1250. {ACM}, 2008.

\bibitem[Dong et~al.(2014)Dong, Gabrilovich, Heitz, Horn, Lao, Murphy,
  Strohmann, Sun, and Zhang]{dong2014knowledge}
Xin Dong, Evgeniy Gabrilovich, Geremy Heitz, Wilko Horn, Ni~Lao, Kevin Murphy,
  Thomas Strohmann, Shaohua Sun, and Wei Zhang.
\newblock Knowledge vault: A web-scale approach to probabilistic knowledge
  fusion.
\newblock In \emph{Proceedings of the 20th ACM SIGKDD international conference
  on Knowledge discovery and data mining}, pages 601--610, 2014.

\bibitem[Bordes et~al.(2013)Bordes, Usunier, Garc{\'{\i}}a{-}Dur{\'{a}}n,
  Weston, and Yakhnenko]{transe}
Antoine Bordes, Nicolas Usunier, Alberto Garc{\'{\i}}a{-}Dur{\'{a}}n, Jason
  Weston, and Oksana Yakhnenko.
\newblock Translating embeddings for modeling multi-relational data.
\newblock In \emph{Advances in Neural Information Processing Systems 26: 27th
  Annual Conference on Neural Information Processing Systems 2013. Proceedings
  of a meeting held December 5-8, 2013, Lake Tahoe, Nevada, United States},
  pages 2787--2795, 2013.

\bibitem[Sun et~al.(2019)Sun, Deng, Nie, and Tang]{rotate}
Zhiqing Sun, Zhi{-}Hong Deng, Jian{-}Yun Nie, and Jian Tang.
\newblock Rotate: Knowledge graph embedding by relational rotation in complex
  space.
\newblock In \emph{7th International Conference on Learning Representations,
  {ICLR} 2019, New Orleans, LA, USA, May 6-9, 2019}, 2019.

\bibitem[Tang et~al.(2020)Tang, Huang, Wang, He, and Zhou]{ote}
Yun Tang, Jing Huang, Guangtao Wang, Xiaodong He, and Bowen Zhou.
\newblock Orthogonal relation transforms with graph context modeling for
  knowledge graph embedding.
\newblock In \emph{Proceedings of the 58th Annual Meeting of the Association
  for Computational Linguistics, {ACL} 2020, Online, July 5-10, 2020}, pages
  2713--2722, 2020.

\bibitem[Zhang et~al.(2019)Zhang, Tay, Yao, and Liu]{quate}
Shuai Zhang, Yi~Tay, Lina Yao, and Qi~Liu.
\newblock Quaternion knowledge graph embeddings.
\newblock In \emph{Advances in Neural Information Processing Systems 32: Annual
  Conference on Neural Information Processing Systems 2019, NeurIPS 2019,
  December 8-14, 2019, Vancouver, BC, Canada}, pages 2731--2741, 2019.

\bibitem[Trouillon et~al.(2016)Trouillon, Welbl, Riedel, Gaussier, and
  Bouchard]{complex}
Th{\'{e}}o Trouillon, Johannes Welbl, Sebastian Riedel, {\'{E}}ric Gaussier,
  and Guillaume Bouchard.
\newblock Complex embeddings for simple link prediction.
\newblock In \emph{Proceedings of the 33nd International Conference on Machine
  Learning, {ICML} 2016, New York City, NY, USA, June 19-24, 2016}, volume~48,
  pages 2071--2080, 2016.

\bibitem[Klicpera et~al.(2019)Klicpera, Bojchevski, and G{\"{u}}nnemann]{appnp}
Johannes Klicpera, Aleksandar Bojchevski, and Stephan G{\"{u}}nnemann.
\newblock Predict then propagate: Graph neural networks meet personalized
  pagerank.
\newblock In \emph{7th International Conference on Learning Representations,
  {ICLR} 2019, New Orleans, LA, USA, May 6-9, 2019}, 2019.

\bibitem[Perozzi et~al.(2014)Perozzi, Al{-}Rfou, and Skiena]{deepwalk}
Bryan Perozzi, Rami Al{-}Rfou, and Steven Skiena.
\newblock Deepwalk: online learning of social representations.
\newblock In \emph{The 20th {ACM} {SIGKDD} International Conference on
  Knowledge Discovery and Data Mining, {KDD} '14, New York, NY, {USA} - August
  24 - 27, 2014}, pages 701--710, 2014.

\bibitem[Hu et~al.(2021)Hu, Fey, Ren, Nakata, Dong, and Leskovec]{hu2021ogblsc}
Weihua Hu, Matthias Fey, Hongyu Ren, Maho Nakata, Yuxiao Dong, and Jure
  Leskovec.
\newblock Ogb-lsc: A large-scale challenge for machine learning on graphs.
\newblock \emph{arXiv preprint arXiv:2103.09430}, 2021.

\end{thebibliography}

\end{document}